\documentclass[journal]{IEEEtran}

\ifCLASSINFOpdf
   \usepackage[pdftex]{graphicx}
 
\else
\fi

\usepackage[cmex10]{amsmath}
\usepackage{array}
\usepackage{tikz}
\usepackage{mdwmath}
\usepackage{mdwtab}
\usepackage{eqparbox}
\usepackage{soul,color}
\usepackage[linesnumbered,ruled,vlined]{algorithm2e}
\usepackage{amssymb}
\newcommand\copyrighttext{%
  \footnotesize \textcopyright 2019 IEEE.Personal use of this material is permitted. Permission from IEEE must be obtained for all other uses, in any current or future media, including reprinting/republishing this material for advertising or promotional purposes,creating new collective works, for resale or redistribution to servers or lists, or reuse of any copyrighted component of this work in other works.}%
\newcommand\copyrightnotice{%
\begin{tikzpicture}[remember picture,overlay]
\node[anchor=south,yshift=10pt] at (current page.south) {\fbox{\parbox{\dimexpr\textwidth-\fboxsep-\fboxrule\relax}{\copyrighttext}}};
\end{tikzpicture}%
}
\begin{document}
\title{Effect of Mutual Coupling on the Performance of STCM-MIMO Systems}
\author{Fatemeh~Asghari~Azhiri,
        Reza~Abdolee,
        and~Behzad~Mozaffari~Tazehkand
\thanks{F.~Asghari~Azhiri and B.~Mozaffari~Tazehkand are with the Department
of Electrical and Computer Engineering, University of Tabriz, Tabriz, Iran e-mail: (f.asghari, mozaffary)@tabrizu.ac.ir}
\thanks{R. Abdolee is with Department of Electrical and Computer Engineering, California State University, Bakersfield, USA e-mail: rabdolee@csub.edu}}
\copyrightnotice
\maketitle
\thispagestyle{empty}
\begin{abstract}
Space-time coded massive (STCM) multiple-input multiple-output (MIMO) system provides superior bit error rate (BER) performance compared with the conventional space-time coding and massive MIMO techniques. The transmitter of the STCM-MIMO system consists of a large antenna array. In a practical system, the self-interference created by the signals transmitted by the elements of this antenna array, known as mutual coupling (MC), degrades the performance of the system. The MC effect is pronounced in communication systems with a large antenna array. On the other hand, increasing the number of transmitting antennas results in improved BER performance. Hence, there is a trade off in selecting the optimum number of transmitting antennas in an STCM-MIMO system. In order to take the impact of MC into account, we have derived an analytical expression for the received signal to accurately model the STCM-MIMO system under the existence of the MC effect. We present an algorithm to select the optimal number of antennas to minimize mutual coupling and the system bit error rate (BER). Through computer simulations, we investigate the BER performance of the STCM-MIMO system for different numbers of array elements. 
\end{abstract}

\IEEEpeerreviewmaketitle
\section{Introduction}
\IEEEPARstart{T}{he} massive multiple-input multiple-output (MIMO) technology has been introduced to meet the growing demands of bandwidth hungry applications and multiuser communications in next generation wireless systems such as fifth generation (5G) cellular networks \cite{bjornson2016massive}. The base station (BS) of a massive MIMO system is able to encode and send data to multiple users simultaneously and in the same frequency band, using a large number of antennas. Therefore, incorporating massive MIMO technology results in robust and highly reliable communications for upcoming ultra dense wireless communication networks \cite{marzetta2016}. 

In order to further improve the performance of massive MIMO systems, Ice et al. \cite{ice2017} introduced ingenious space-time coded massive MIMO (STCM-MIMO) technique that benefits from the advantages of both massive MIMO and space-time coding technologies. The data symbols in a STCM-MIMO system are coded by a space-time code technique in the transmitter. Then, each of the coded symbols is transmitted by a subset of the antenna array in the BS.  Using antenna arrays in STCM-MIMO for transmitting the space-time coded symbols significantly improves the overall system performance compared to the space-time coded MIMO and conventional massive MIMO system \cite{ice2018}.

The transmitter of the STCM-MIMO system contains a large number of antennas that must be located close to each other in order to achieve an appropriate size of the base station. This compactness leads to an increase in the mutual coupling (MC) effect between elements of the antenna arrays.

The effect of the MC should be considered in the design and implementation of the communication systems specifically massive MIMO based systems due to the deployment of large antenna arrays in their structure. The impact of MC and the spatial correlation have been studied on various types of systems with array antennas in their structure. The effect of MC in BER performance of Alamouti space-time coded systems has been investigated in \cite{abouda2006effect} and a performance degradation in low correlated channels has been indicated. In \cite{RaoBERAO} an antenna selection algorithm is used in the receiver side in the presence of mutual coupling to achieve best BER performance for an Alamouti coding scheme. The authors in \cite{wallace2004mutual} proposed a rigorous network-theory framework for the analysis of mutual coupling in MIMO wireless communications. This method attains an upper bound for the capacity expression in the presence of mutual coupling in the studied system.

The growing application of massive MIMO technology in the new generations of communication networks such as 5G systems has attracted increasing attention to study its hardware performance. In \cite{artiga2012mutual}, signal to noise and interference ratio of a massive MIMO system for various types of antenna arrays are calculated. The mutual coupling model of antenna arrays is applied to the 3GPP 3D channel model in \cite{pratschner2017mutual} and by a matching network, the coupling effect has been partly compensated. Most articles in the literature assume that the antenna arrays have a regular structure such as uniform linear or uniform planar arrays. The antenna arrays may have irregular structures as shown in \cite{ge2016multi}. It is indicated that in some specific cases, the irregular arrays outperform the regular antenna arrays in the achievable rate.

Since STCM-MIMO systems are introduced recently, their hardware implementation has not been investigated thoroughly. It has been shown that increasing the number of antenna elements in the transmitter of an STCM-MIMO system results in better BER performance, while the destructive effect of MC of the antenna array is not considered \cite{ice2018}. However, the effect of MC is not negligible in practice, especially for large antenna arrays. In this paper, we study the performance of STCM-MIMO system in the presence of the MC effect. We derive the analytical expression of the received signal vector by considering the MC effect of the antenna elements in the transmitter. The BER performance of an STCM-MIMO system which consists of a uniform linear array (ULA) antenna is investigated.
The simulation results determined the amount of the performance degradation of the STCM-MIMO system with different number of antenna elements and element distances in the presence of the coupling effect. In order to design a proper array structure for the transmitter of an STCM-MIMO system, we compare the performance of the system with various number of antennas within identically sized arrays and different styles of sub-array selection in the transmitter. 
The simulation results confirm that we face a trade off in specifying the optimum number of elements of the antenna array to achieve appropriate system performance. Even though increasing the number of antennas in STCM-MIMO transmitter improves the BER in ideal systems, it ends in increasing the destructive effect of mutual coupling when the MC effect is taken into account. We investigate this trade off and determine the optimum number of elements for the inspected antenna arrays.

The remainder of the paper is organized as follows. In \hbox{section \ref{sec2}} the system model is discussed and the received signal vector is formulated. The simulation results are discussed in section \ref{sec3} followed by concluding remarks in section \ref{sec4}.

\section{System Model}\label{sec2}
\begin{figure}[!t]
\centering
\includegraphics[width=3.4 in]{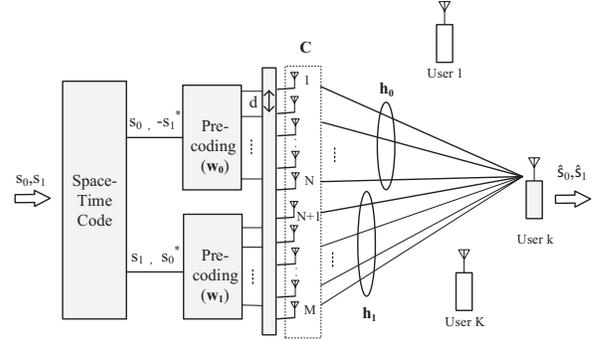}
\caption{$2N \times 1$ STCM-MIMO}
\label{sysmod}
\end{figure}

We consider an STCM-MIMO system with a base station equipped with $M$ antennas and $K$ single antenna mobile users. We divide the base station array into two sub-arrays each having $N=M/2$ elements. Two information symbols with a certain modulation scheme such as M-ary QAM or PSK modulation are transmitted by two transmit antenna sub-arrays in each time slot. This system applies the Alamouti space-time codes and uses Hermitian pre-coding scheme \cite{ice2017}. At time $t$, the first sub-array sends the symbol $s_{0}$ and the second sub-array sends $s_{1}$. At time $t+T$, the symbols $-s_{1}^{*}$ and $s_{0}^{*}$ are sent by first and second sub-arrays, respectively. The Hermitian pre-coding is applied on each symbol before transmitting (see Figure \ref{sysmod}). The weight vector $w_{0}$ ($w_{1}$) is the Hermitian pre-coding vector according to the channel vector of the first (second) sub-array and the receiver.

The received signals can be expressed as \cite{ice2017} 
\begin{equation}\label{eq1}
\begin{split}
\tilde{r}_{0}^{nc}=&\tilde{r}^{nc}(t)=\boldsymbol{w}_{0}^{H}\boldsymbol{h}_{0}s_{0}+\boldsymbol{w}_{1}^{H}\boldsymbol{h}_{1}s_{1}\\
&+\sum_{j\neq0}^{K-1}(\boldsymbol{w}_{2j}^{H}\boldsymbol{h}_{0}s_{2j}+\boldsymbol{w}_{(2j+1)}^{H}\boldsymbol{h}_{1}s_{(2j+1)})+\tilde{n_{0}}
\\
\\
\tilde{r}_{1}^{nc}=&\tilde{r}^{nc}(t+T)=-\boldsymbol{w}_{0}^{H}\boldsymbol{h}_{0}s_{1}^{*}+\boldsymbol{w}_{1}^{H}\boldsymbol{h}_{1}s_{0}^{*}\\
&+\sum_{j\neq0}^{K-1}(-\boldsymbol{w}_{2j}^{H}\boldsymbol{h}_{0}s_{2j+1}^{*}+\boldsymbol{w}_{(2j+1)}^{H}\boldsymbol{h}_{1}s_{(2j)}^{*})+\tilde{n_{1}}
\end{split}
\end{equation}
where $\tilde{r}_{0}^{nc}$ is the received signal at time slot $t$ without MC effect, $\tilde{r}_{1}^{nc}$ is the received signal at time slot $t+T$ without MC effect, $K$ is the number of users, $(.)^{H}$ and $(.)^{*}$ indicate Hermitian transform and complex conjugate respectively. The coefficient vector of the channel between the antenna array at the transmitter and the single antenna receiver is expressed as $\boldsymbol{h}=[\boldsymbol{h}_{0}; \boldsymbol{h}_{1}]$, where $\boldsymbol{h}_{0}$ and $\boldsymbol{h}_{1}$ are $N \times 1$ vectors which exhibit the channel coefficients through first and second sub-arrays to the receiver, respectively. The massive MIMO Hermitian pre-coding vector, $\boldsymbol{w}_{j}$, is  defined as
\begin{equation}
\boldsymbol{w}_{j}=\frac{1}{N}\boldsymbol{h}_{j}
\end{equation}

The receiver combining scheme for two branches of STCM-MIMO system can be expressed as
\begin{equation}\label{eqr}
\begin{split}
\tilde{s}_{0}&=\Vert \boldsymbol{h}_{0}\Vert^{2}\tilde{r}^{nc}_{0}+\Vert \boldsymbol{h}_{1}\Vert^{2}\tilde{r}^{nc*}_{1}\\
\tilde{s}_{1}&=\Vert \boldsymbol{h}_{1}\Vert^{2}\tilde{r}_{0}^{nc}-\Vert \boldsymbol{h}_{0}\Vert^{2}\tilde{r}_{1}^{nc*}
\end{split}
\end{equation}
where $\Vert\boldsymbol{v}\Vert ^{2}$ is the $L^{2}$-norm of the vector $\boldsymbol{v}$. Using the equation (\ref{eqr}) the maximum likelihood detector can estimate the transmitted symbols.

Equation (\ref{eq1}) assumes that the antenna elements of the array do not interfere with each other. When antenna elements are close, the electromagnetic field produced by one antenna influences the output of its neighbor antennas. The interaction between two or more antennas, that affects coefficients of the antenna array is called mutual coupling.

Assuming that a communication system consists of an antenna array with $M$ elements as a transmitter and a single antenna receiver, the received signal at the receiver can be written as
\begin{equation}\label{eq3}
y^{nc}=\boldsymbol{gx}
\end{equation}
where $y^{nc}$ is the received signal without MC effect, $\boldsymbol{x}$ is the transmitting signal vector and $\boldsymbol{g}$ is the wireless channel vector. 

Mutual coupling effect can be incorporated into (\ref{eq3}) as follows
\begin{equation}
y^{c}=\boldsymbol{Cgx}=\hat{\boldsymbol{g}}\boldsymbol{x}
\end{equation}
where $\boldsymbol{C}$ represents the $M \times M$ coupling matrix and $\boldsymbol{\hat{g}}$ can be replaced as the channel vector in order to consider MC effect.

Therefore, in the case of STCM-MIMO, the receiving signal including mutual coupling can be written as
\begin{equation}\label{eq6}
\begin{split}
\tilde{r}_{0}^{c}=&\tilde{r}^{c}(t)=\boldsymbol{w}_{0}^{H} \hat{\boldsymbol{h}}_{0}s_{0}+\boldsymbol{w}_{1}^{H} \hat{\boldsymbol{h}}_{1}s_{1}\\
&+\sum_{j\neq0}^{K-1}(\boldsymbol{w}_{2j}^{H} \hat{\boldsymbol{h}}_{0}s_{2j}+\boldsymbol{w}_{(2j+1)}^{H} \hat{\boldsymbol{h}}_{1}s_{(2j+1)})+\tilde{n_{0}}
\\
\\
\tilde{r}_{1}^{c}=&\tilde{r}^{c}(t+T)=-\boldsymbol{w}_{0}^{H} \hat{\boldsymbol{h}}_{0}s_{1}^{*}+\boldsymbol{w}_{1}^{H} \hat{\boldsymbol{h}}_{1}s_{0}^{*}\\
&+\sum_{j\neq0}^{K-1}(-\boldsymbol{w}_{2j}^{H} \hat{\boldsymbol{h}}_{0}s_{2j+1}^{*}+\boldsymbol{w}_{(2j+1)}^{H} \hat{\boldsymbol{h}}_{1}s_{(2j)}^{*})+\tilde{n_{1}}
\end{split}
\end{equation}
where the wireless channel between the transmitter array and the receiver of the system including the MC effect is defined as
\begin{equation}
\hat{\boldsymbol{h}}= \boldsymbol{Ch}= \boldsymbol{C} 
\begin{bmatrix}
\boldsymbol{h}_{0} \\
\boldsymbol{h}_{1}
\end{bmatrix}=
\begin{bmatrix}
\hat{\boldsymbol{h}}_{0} \\
\hat{\boldsymbol{h}}_{1}
\end{bmatrix}
\end{equation}

The mutual coupling matrix, $\boldsymbol{C}$ can be acquired from electromagnetics analysis and measurement. The antennas have the reciprocity property which means that the receive and transmit properties of an antenna are identical \cite{Balanis}. Therefore, in the case that the receiver uses an antenna array, the mutual coupling effect will be the same.

For a uniform linear array consisting of $M$ dipole elements, $\boldsymbol{C}$ can be written as \cite{cuiwei2016effect}
\begin{equation}\label{eq8}
\boldsymbol{C}=(Z_{A}+Z_{L})(\boldsymbol{Z}+Z_{L}\boldsymbol{I})^{-1}
\end{equation} 
where $\boldsymbol{I}$ is the identity matrix of size $M \times M$ and $Z_{A}$ and $Z_{L}$ are the antenna impedance and load impedance, respectively. The elements of matrix $\boldsymbol{Z}$ can be calculated as follows \cite{masouros2013large}
\begin{equation}
Z_{mn}=\left\lbrace \begin{array}{ll}
\begin{split}
\frac{\eta_{0}}{4\pi}&[0.577+ln(2\pi)-Ci(2\pi)\\
&+jSi(2\pi)]\end{split} & m=n\\

\begin{split}
\frac{\eta_{0}}{4\pi}&\{[2Ci(\beta d)-Ci(\beta u_{1})-Ci(\beta u_{2})]\\
&-j[2Si(\beta d)-Si(\beta u_{1})-Si(\beta u_{2})]\} \end{split} & m \neq n
\end{array}\right. 
\end{equation}
where $\eta_{0}= \sqrt{\mu_{0}/ \epsilon_{0}}\approx 120\pi$ is the intrinsic impedance and $\beta=2\pi/\lambda$ is the wave number, $\lambda$ is the wavelength and
\begin{equation}
\left\lbrace \begin{array}{ll}
u_{1}=\sqrt{d^{2}+L^{2}}+L \\
u_{2}=\sqrt{d^{2}+L^{2}}-L
\end{array}\right. 
\end{equation}
where $d$ is the distance between array elements and $L$ is the length of the dipole antenna. $Ci(x)$ and $Si(x)$ are the cosine and sine integrals defined as
\begin{equation}\begin{split}
&Ci(x)=\int^{x}_{- \infty} \frac{\cos(x)}{x} dx \\
&Si(x)=\int^{x}_{- \infty} \frac{\sin(x)}{x} dx
\end{split}
\end{equation} 
\section{Computer Experiment Results}\label{sec3}
In order to investigate the effect of MC on the performance of STCM-MIMO systems, it is essential to calculate the coupling matrix which depends on the array structure. In this section, we analyze an STCM-MIMO communication system with a ULA as its transmitter and following assumptions and parameters. A $2N \times 1$ STCM-MIMO system with one base station is assumed which contains a transmitting antenna array with $M$ elements. The antenna array is a ULA consisting of dipole antennas with length $L$ and element spacing $d$. The carrier frequency is assumed to be 2 GHz. Data symbols are chosen from QPSK modulation constellation. The full channel state information (CSI) is available at the transmitter for calculating pre-coding weights and at the receiver for decoding the data. The STCM-MIMO system utilizes $2 \times 1$ Alamouti code, so two consecutive time slots are required for detecting two transmitted symbols. The transmission environment is assumed to be dispersive such that the channel coefficients follow i.i.d. complex Gaussian distribution; also, there exists additive white complex Gaussian noise at the receiver. All the results are achieved by 20000 times Monte-Carlo simulations.

\begin{figure}[!t]
\centering
\includegraphics[width=3 in]{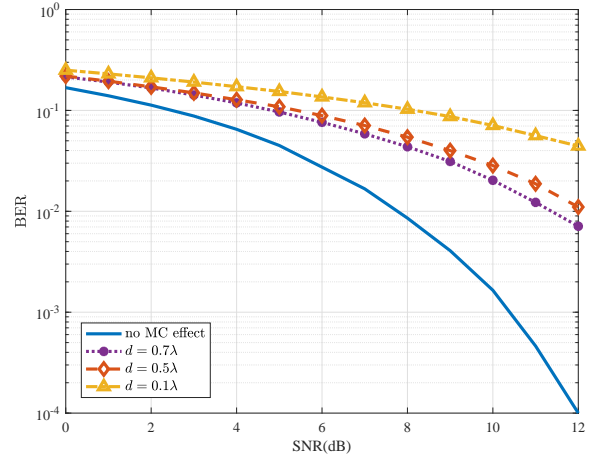}
\caption{BER performance of STCM-MIMO with MC effect, $M=100$}
\label{fig2}
\end{figure}

Figure \ref{fig2} represents the MC effect on the BER performance of STCM-MIMO system. In this simulation, the transmitting array consists of 100 antenna elements with various distances. The antenna impedance is $Z_{A}=73+42j$ and the load impedance is assumed to be $Z_{L}=Z_{A}^{*}$ in order to provide full matching. The simulation results show that, decreasing the distances between array elements increases the MC effect between array antennas. Hence, the bit error rate increases for the same SNR value.

\begin{figure}[!t]
\centering
\includegraphics[width=2.87 in]{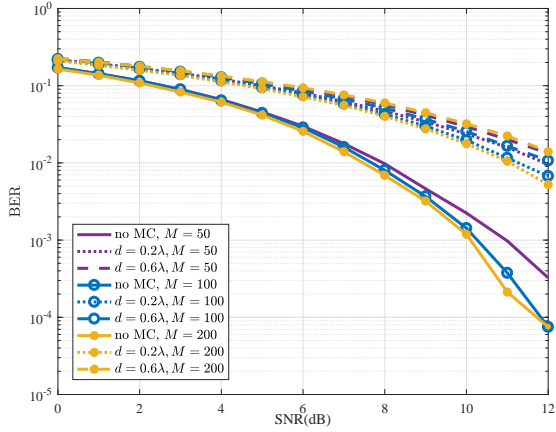}
\caption{BER performance of STCM-MIMO with MC effect for different number of array elements, $d=0.2\lambda,d=0.6\lambda$}
\label{fig3}
\end{figure}

In a STCM-MIMO system as long as the MC effect of the antenna elements is negligible, it's straightforward to increase the number of elements of the antenna array in the transmitter to improve the BER performance. However, by increasing the number of elements in the antenna array of the transmitter or decreasing the distances between them, the MC effect become significant. Figure \ref{fig3} demonstrates the BER performances of the STCM-MIMO systems with different numbers of antenna elements in the presence of MC effect. 

\begin{figure}[!t]
\centering
\includegraphics[width=3 in]{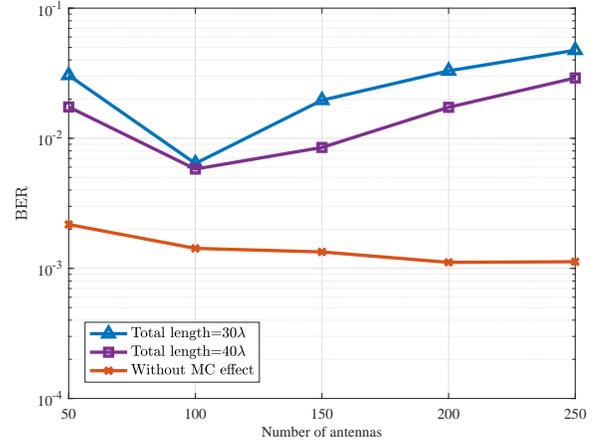}
\caption{BER performance of 2$\times$1 STCM-MIMO versus total number of antennas determined for fixed array size at SNR=10 dB}
\label{fig4}
\end{figure}

\begin{figure}[!t]
\centering
\includegraphics[width=3 in]{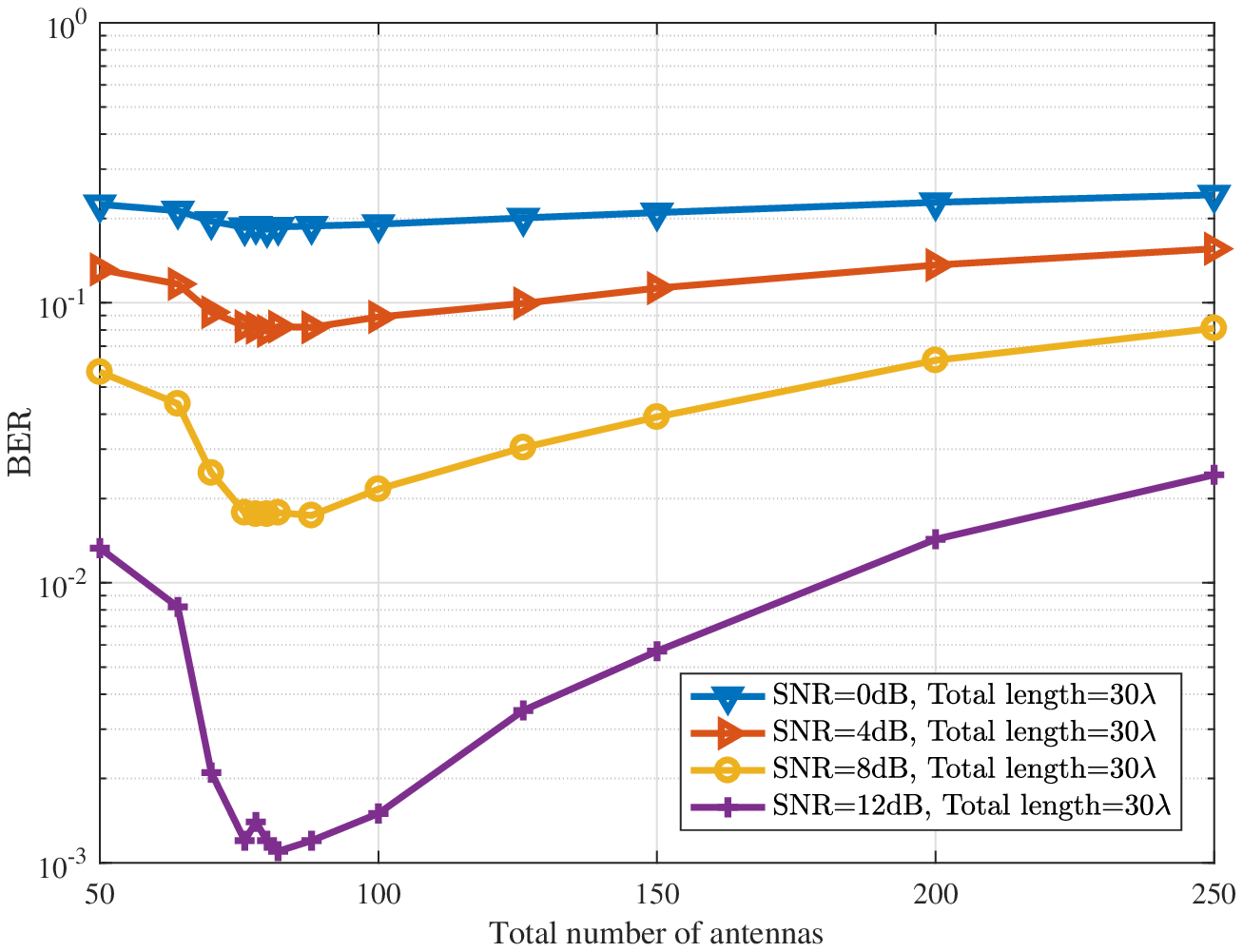}
\caption{BER performance of STCM-MIMO with MC effect versus total
number of antennas determined for array size of $30\lambda$ at different SNR
conditions}
\label{fig6}
\end{figure}

\begin{figure}[!t]
\centering
\includegraphics[width=3 in]{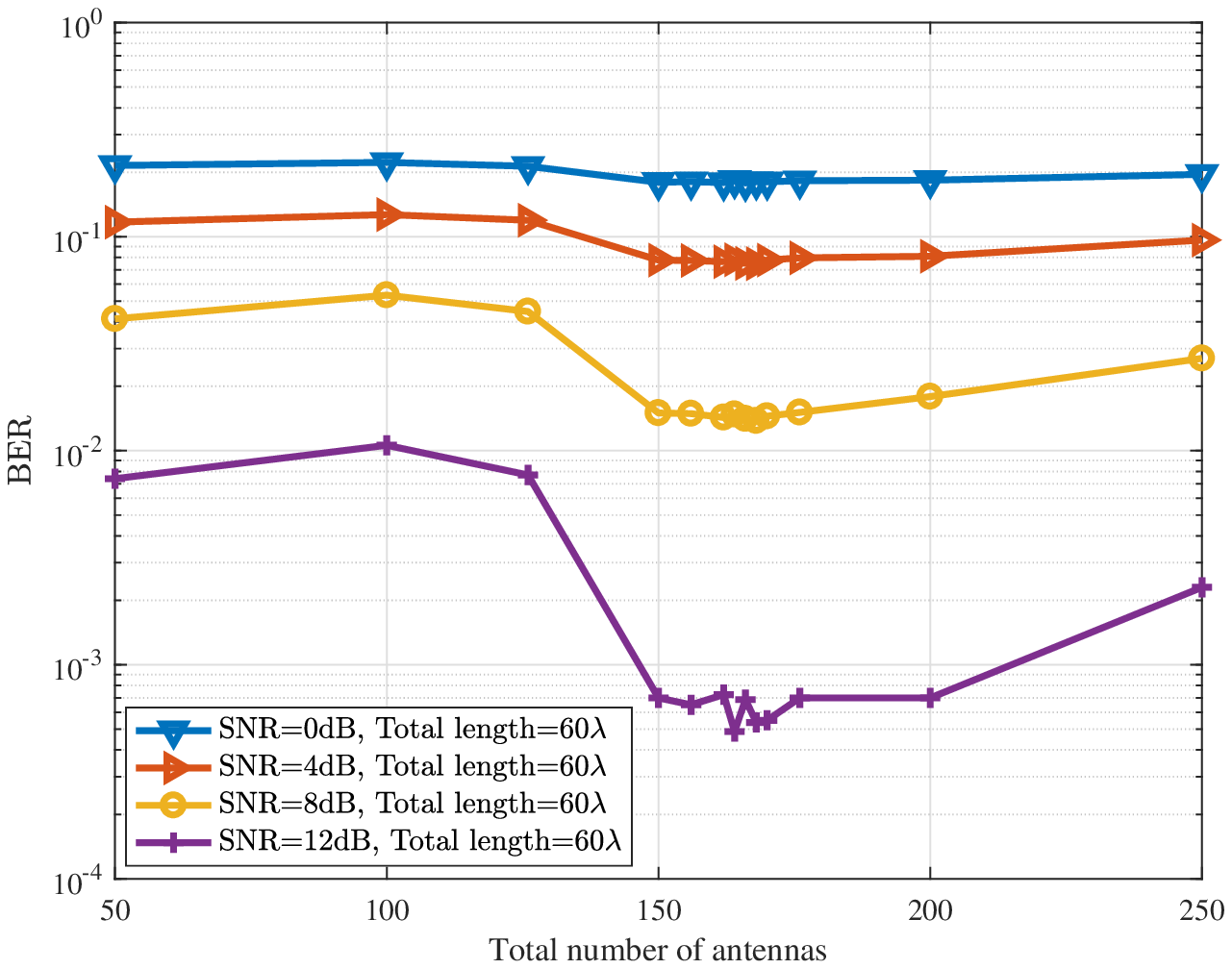}
\caption{BER performance of STCM-MIMO with MC effect versus total
number of antennas determined for array size of $60\lambda$ at different SNR
conditions}
\label{fig7}
\end{figure}

To achieve the optimum performance in a system with compact transmitter, it is important to utilize the appropriate number of antenna elements with specific distances in its base station. In a STCM-MIMO system without considering the MC effect, increasing the number of transmitting antennas improves the BER performance {\cite{ice2018}}. Figure {\ref{fig4}} compares the BER performance of a STCM-MIMO system with and without considering the MC effect at an SNR of 10dB. Our simulation results demonstrate a trade-off to select the optimum number of antenna elements. Increasing the number of elements may improve the performance of STCM-MIMO, however to keep the size of the transmitter constant, the distances between elements have to be decreased, which increases MC effect and degrades BER performance. The antenna array of the transmitter is considered to be a uniform linear array consisting of $M$ elements with uniform spacing. The analytical formulation of mutual coupling effect presented in {\eqref{eq6}} facilitates determining the optimal number of antennas in a STCM MIMO system. We propose a  heuristic search algorithm inspired by Newton-Raphson method to achieve the optimum number of antennas (Algorithm {\ref{alg}}). We first divide the searching interval by the arbitrarily chosen 'step' size which gives us sample numbers of antennas, $n_{i}$. The sample numbers must be even numbers in a $2 \times 1$ STCM-MIMO system. Then we calculate the MC matrix for $n_{i}$s using {\eqref{eq8}}. In the next step, the BER of the STCM-MIMO system is calculated utilizing the {\eqref{eq6}} and by simulation for the intended SNR values. Then $n_{j}=\arg \min(average(BER)), n_{1}\leq n_{j}\leq n_{m}$ is selected. The searching is continued among $n_{j}$, $2\lceil \frac{n_{j-1}+n_{j}}{4} \rceil$ and $2\lceil\frac{n_{j}+n_{j+1}}{4}\rceil$ where $\lceil.\rceil$ denotes the ceil function. The number with $\min(average(BER))$ is selected to be the middle sample and the searching continues in the adjacency of this number. The rest of searching process continues until reaching the optimal number of antennas with minimum $average(BER)$. Figures {\ref{fig6}} and {\ref{fig7}} show the BER performance of STCM-MIMO system for $total \: length=30\lambda$ and $60\lambda$ at various SNR values. In Table {\ref{table}} we provide the optimum number of transmitting antennas in a $2\times 1$ STCM-MIMO system for the antenna arrays with specified total lengths. The required information for the search algorithm have been considered as $search \: interval=(50,250)$ and $step=50$.


\begin{figure}[!t]
\centering
\includegraphics[width=3 in]{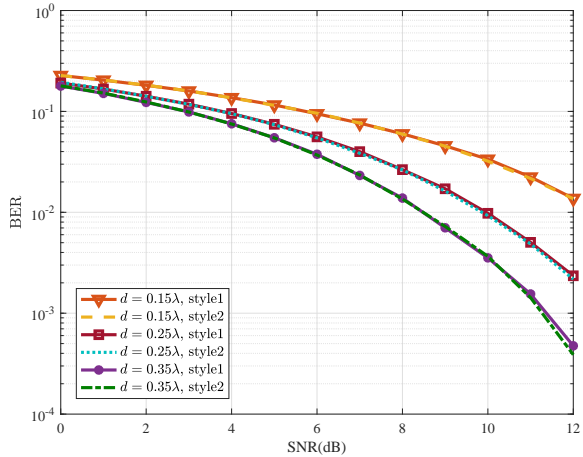}
\caption{Antenna selection effect on BER performance, M=200}
\label{fig5}
\end{figure}
In the previous simulations, the antenna sub-arrays of first and second branches have been selected as shown in Figure~\ref{sysmod} (style 1). Simulation results don't show significant performance differences with the case that the sub-arrays were chosen one in between (style 2). Figure \ref{fig5} demonstrates the results of this comparison for an antenna array with 200 elements. 
\begin{table}[t]
\centering
\caption{Optimal number of antennas in the transmitter of STCM-MIMO systems}\label{table}
\begin{tabular}{|c|c|}
\hline 
Total length of array & Optimum number of antennas \\ 
\hline 
$30\lambda$ & 80 \\ 
\hline 
$40\lambda$ & 112 \\ 
\hline 
$50\lambda$ & 136 \\ 
\hline 
$60\lambda$ & 166 \\ 
\hline 
$70\lambda$ & 188 \\ 
\hline 
\end{tabular}
\end{table} 

\begin{algorithm}[t]
\caption{Search Algorithm to find the optimal number of antennas in the transmitter of a 2$\times$1 STCM-MIMO system}\label{alg}
\KwIn{Search interval: ($n_{1}$, $n_{m}$), step, total length}
\KwOut{$n_{j}$}
\While{$\min (n_{j}-n_{j-1},n_{j+1}-n_{j})>2$}{
    Calculate \textbf{$C_{i}$} for $M=n_{i}$ and $d=\dfrac{total \: length}{n_{i}}$ via \eqref{eq8}, 
    where $n_{i}=2\lceil \frac{n_{1}+k.step}{2}\rceil$, $k=0,1,2,\ldots$ , $n_{1}\leq n_{i}\leq n_{m}$. \\
    Calculate BER using received signals of \eqref{eq6} \\
    $n_{j}=\arg \min(average(BER))$\\
    $n_{j-1} \gets 2\lceil \frac{n_{j-1}+n_{j}}{4} \rceil$ \\
    $n_{j+1} \gets 2\lceil\frac{n_{j}+n_{j+1}}{4}\rceil$ \\
}
\end{algorithm}

\section{Conclusion}\label{sec4}
In this paper, the STCM-MIMO system has been modeled considering the MC effect using an analytical expression derived for the received signal of the system. The simulation results indicated a performance degradation in the presence of MC. The existence of a trade-off has been demonstrated for determining the appropriate number of antennas in the transmitters with predefined physical sizes. Increasing the number of array elements improves the BER performance due to the increment of the number of antennas in STCM-MIMO transmitter, on the other hand it increases the destructive effect of mutual coupling and causes degradation in system performance. We proposed an algorithm to calculate the optimal number of antennas to achieve the minimum bit error rate of STCM-MIMO systems.

\ifCLASSOPTIONcaptionsoff
  \newpage
\fi

\bibliographystyle{IEEEtran}
\bibliography{ref.bib}









\end{document}